\documentclass[a4paper]{aa}

\usepackage{psfig}
\usepackage{times}

\def\simlt {\lower.5ex\hbox{$\; \buildrel < \over \sim \;$}}
\def\simgt{\lower.5ex\hbox{$\; \buildrel > \over \sim \;$}}   
\def\msun {\hbox{M$_{\odot}$}} 
\def\Te {$T_{\rm e}$}
\def\net {$n_{\rm e}t$\ }

\def\EM {$n_{\rm e} n_{\rm H} V/d^2$}   
\def\NH {$N_{\rm H}$\ }

\begin{document}
   \thesaurus{ 09.09.1; 09.19.2; 13.25.4}

    \title{The BeppoSAX X-ray spectrum of the remnant of SN 1006}

    \author{Jacco Vink\inst{1}\,\inst{2} \and Jelle S.  Kaastra\inst{1} \and Johan A. M. Bleeker\inst{1} \and Andrea Preite-Martinez\inst{3}}

    \offprints{J. Vink (jvink@aip.de)}

    \institute{
	SRON, Laboratory for Space Research,
	Sorbonnelaan 2, NL-3584
        CA Utrecht, The Nether\-lands 
	\and
	Present address: Astrophysikalisches Institut Potsdam,
	An der Sternwarte 16, D-14482 Potsdam, Germany
	\and
	Istituto di Astrofisica Spaziale,
	Area di Ricerca di Tor Vergata,
	Via del Fosso del Cavaliere 00133,
	Roma, Italy
	}
\date{Received  /  Accepted }

\maketitle

\begin{abstract}
We present BeppoSAX X-ray spectra of the remnant of the supernova of AD 1006,
which cover a broad spectral range of 0.1-10~keV.
In our analysis we concentrate on the thermal emission from
the central region of the remnant.
For this purpose we fitted the spectra of the whole remnant 
using a new version of the spectral program SPEX, 
which takes into account the interdependence of spectra of different regions 
as the result of the point spread function of the instruments.
This is in particular important for separating the synchrotron emission,
coming from the bright rims of the remnant, 
from the thermal emission from the central regions.

The thermal emission appears best fitted with a combination of two spectral
components with electron temperatures of $\sim$0.7~keV and $\simgt$3~keV.
The plasma in SN 1006 appears to be very far out of ionization equilibrium
(\net $\sim 2\ 10^{15}$~m$^{-3}$s), which means that the L-shell
emission of magnesium, silicon and sulphur contribute significantly to
the flux below 0.5~keV.
We also note the presence of Fe K line emission at $6.3\pm0.2$~keV,
in agreement with an origin from inner shell ionizations and excitations.
We confirm that the abundances in SN 1006 are clearly non-solar.
In particular silicon is very abundant.
We did not find substantial variations in plasma properties over the face of 
the remnant, although the hotter component seems more dominant in the Northern
half of the remnant.
\keywords{
ISM: individual objects: SN 1006 -- ISM: supernova remnants -- X-rays: ISM }  
\end{abstract}

\section{Introduction}
The historical supernova SN 1006 was probably a Type Ia supernova
(Clark \& Stephenson \cite{Clark}, Schaefer \cite{Schaefer}). 
Its remnant has now a diameter of 30' and in recent years three 
interesting findings were reported:
\begin{itemize}
\item
The ASCA X-ray spectrum indicates that 
the emission above $\sim 1$~keV is dominated by synchrotron
radiation from electrons with energies up to $\sim$100~TeV, 
which are accelerated at the shock front
(Koyama et al. \cite{Koyama95}, Reynolds \cite{Reynolds98}). 
The detection of TeV gamma ray emission confirms the presence of extremely 
relativistic electrons (Tanimori et al. \cite{Tanimori}).\\
\item 
Modeling far ultra-violet spectra of a Northwestern filament 
has revealed that the post shock plasma has not reached 
ion-electron temperature equilibration. 
The electron temperature was found to be only 
$10\%$ of the ion temperature (Laming et al. \cite{Laming96}).  \\
\item 
Ultraviolet absorption features in the spectrum of the 
Schweizer-Middleditch (\cite{Schweizer}) star, 
a blue subdwarf lying  behind the remnant,
have revealed the presence of blue and red-shifted unshocked silicon and iron
(Hamilton et al. \cite{Hamilton97}). 
Remarkably, no evidence was found for the presence of $\sim 0.5$\,\msun\ of 
iron, which is likely to have been synthesized during the carbon deflagration 
of the white dwarf progenitor (Nomoto et al. \cite{Nomoto}). 
Carbon deflagration is the currently favored model 
for Type Ia supernovae. 
\end{itemize}

The presence of a synchrotron component contaminates the thermal X-ray
emission and
the non-equilibration of ion and electron temperatures 
obscures the underlying shock
hydrodynamics, which might otherwise have been inferred from the temperature 
structure.
On the other hand, in X-rays other ion species can be studied than in the UV 
and this may help verifying the models for nucleosynthesis of Type Ia
supernovae.

In this paper we present BeppoSAX spectra of the remnant of SN 1006.
We will show that the thermal emission from the central region of the remnant 
is best described by a two component non-equilibrium ionization model with
a very low ionization parameter. 

\begin{figure*}[th]
\vskip 8.2truecm
	\caption{The right images shows the sky regions for the model spectra.
 The image on the left shows a ROSAT HRI mosaic of SN 1006 on the same scale.
\label{sectors}}
\end{figure*}

\section{The data and method}
The Italian-Dutch satellite BeppoSAX (Boella et al. \cite{Boella97a}) observed 
SN 1006 in April 1997. The observation consists of three different pointings, 
which together cover the whole remnant by the imaging instruments 
LECS (Parmar et al. \cite{Parmar}) and MECS (Boella et al. \cite{Boella97b}). 
The total observation time is 91~ks for the MECS and 37~ks 
for LECS,
which has a lower observation efficiency owing to UV leakage. 

Both the LECS ($\sim 0.1 - 10$~keV) 
and the MECS ($\sim 2-10$~keV) are gas scintillation counters with a 
spectral resolution of $\sim 8$\% at 6~keV, a spatial resolution at 6~keV
of $\sim 2$\arcmin\ FWHM and a half power width\footnote{The half power width gives the width of the area in which 50\% of the photons of a point  source are expected to end up; this measure is sensitive to the broad wings of the point spread function, whereas the FWHM is more determined by the core of the point spread function.}
of $\sim 2.5$\arcmin.
The spatial resolution is strongly energy dependent: with increasing 
energy the core of the point spread function decreases, 
whereas the wings of the point spread function 
(determined mostly by the mirror properties) become more dominant
(Parmar et al. \cite{Parmar}, Boella et al. \cite{Boella97b}, 
cf. Vink et al. \cite{Vink99}).
The extended wings of the point spread function make that in the case of 
SN 1006 the spectra of the low brightness central region,
which emits mostly thermal emission
(Koyama et al. \cite{Koyama95}, Willingale et al. \cite{Willingale}, 
the latter based on ROSAT PSPC data), 
is contaminated by the radiation from the rims of the
remnant, which is predominantly synchrotron radiation.

We circumvented this problem by using a new
version (v2.0) of the X-ray spectral fitting program SPEX 
(Kaastra et al.~\cite{Kaastra}, cf. Kaastra et al. \cite{Kaastra99}), 
with which one can fit several spectra from the same object simultaneously.
In this version the concept of the spectral redistribution matrix 
is extended to include also a spatial redistribution part. 
The spectral model is calculated on a spatial/spectral input grid 
and a model of the 
instrument properties describes what fraction of the incoming
photons with a certain energy, coming from a certain region of the sky
will end up in a given energy bin (channel) and spatial bin 
(spectral extraction region).
This means in practice that two sets of spatial regions have to be provided
(in the program these are specified using images compliant 
with the FITS format). 
One set specifies the sectors for which the model is calculated, 
the other set specifies the spectral extraction regions. 
For stable calculations, the extraction regions should roughly
correspond to the spatial model sectors, but a detailed correspondence 
is not necessary. 
For example, some regions of SN 1006 were only partially covered by some
observations, but this is accounted for in the spectral/spatial 
redistribution matrix.
The X-ray emission models used for this paper are the same as in the 
standard SPEX program.

We divided the remnant into six regions. Apart from a central region and 
two X-ray bright rims, we divided these regions further into a 
Northern and Southern half in order to see if the thermal emission from
the brighter Southern half is different from the emission from the 
Northern halve. With North and South we mean here, and throughout the rest of 
the text, 
North and South with respect to the principle axis of the remnant, 
which is tilted with respect
to the North-South direction in equatorial coordinates.
The spatial model grid for which the models were calculated are shown in 
Fig.~\ref{sectors}.
We chose spectral extraction regions roughly corresponding to the model 
regions,
but the extraction regions corresponding to the bright rims were larger,
in order to reduce the contamination of the central regions 
by emission coming from the bright rims.
We used archival ROSAT HRI images (Fig.~\ref{sectors}, see also 
Winkler \& Long \cite{Winkler}) to
define the spectral extraction regions and the model sectors.
For the spectral extraction regions we took also into account the extend 
of the synchrotron rims in the BeppoSAX images, as they seem broader than on
the HRI images due to  the point spread function.

The extraction of the spectra and the generation of the spectral-spatial
response matrix was done with a program specifically made to
extract BeppoSAX spectra. 
It incorporates the current knowledge of the
detectors, such as the point spread function, vignetting and absorption by the
strong-backs. We used the response matrices of September 1997 and 
the standard background data (November 1998).
However, there are still some uncertainties in the detector 
calibration, especially for off axis positions. 
These uncertainties are larger for the LECS, which has a more complicated 
design than the MECS. For this reason we left the normalization of the 
LECS spectra and some off-axis MECS spectra free with respect to the 
on-axis MECS spectra. 
The relative LECS normalizations turned out to be between 0.6 to 0.9, 
consistent with other BeppoSAX results (e.g. Favata et al. \cite{Favata}).
The spectra were binned to a bin size of roughly 1/3rd the spectral resolution
and some further rebinning was done for channels with low count rates.
In order to circumvent statistical problems with bins with very few counts
we used a method proposed by Wheaton et al. (\cite{Wheaton}). 
This means that after obtaining a good fit, we used the best fit model to 
calculate the expected error per bin, instead of the observed counts.
Using this method with two or three extra iterations gives in general a
stable and in principle more reliable $\chi^2$ value.

\begin{figure}
	\psfig{figure=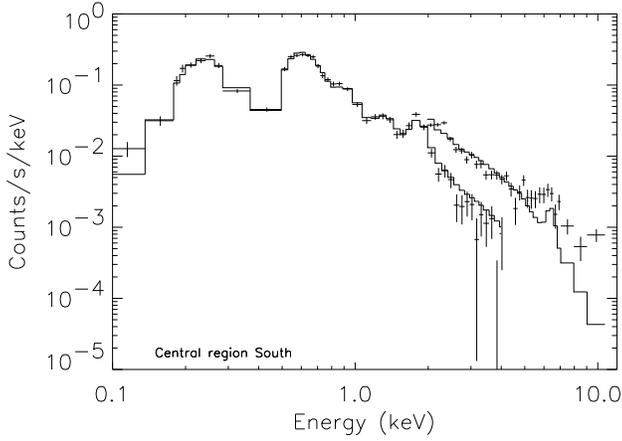,width=8.5cm}
	\caption{The combined LECS/MECS spectrum of the Southern central region, with a simple model fit (i.e. no hard thermal or non-thermal component),
which does not adequately fit the spectrum above $\sim 5$~keV.
\label{hard_excess}}
\end{figure}

\begin{table}
        \caption{
The best fitting parameters for our model with two NEI thermal components.
The power law components for the emission from the rims are listed separately
in Table~\ref{powerlaw}.
The abundances are relative to solar abundances 
(Anders \& Grevese \cite{Anders}). Errors are 90\% confidence limits and
lower/upper limits are 99\% confidence limits.
\label{parameters}}
	\begin{flushleft}          
                \begin{tabular}{l ll}
                        \hline\noalign{\smallskip}
 & North & South\\
                        \hline\noalign{\smallskip}
\EM\ I ($10^{62}$m$^{-3}$kpc$^{-2}$) & $2.3 \pm 1.0$ & $3.9\pm 1.0$\\
\Te\  I (keV)                 & $0.71 \pm 0.15$ & $0.78 \pm 0.09$ \\
\net\ I ($10^{15}$~m$^{-3}$s) & 3.3 ($<4.6$)    & 2.3 ($<3.3$)\\
\noalign{\smallskip}
\EM\  II ($10^{62}$m$^{-3}$kpc$^{-2}$) &  $0.1\pm 0.04$   & $0.14 \pm 0.03$  \\
\Te\  II (keV)                 & 3.4 ($> 3$) & 8.0 ($> 4$) \\
\net\ II ($10^{15}$~m$^{-3}$s) & 0.7 ($< 2$) & 0.7 ($<$3 ) \\
\noalign{\smallskip}
C   &     \multicolumn{2}{c}{$0.8\pm0.4$} \\
O   &  \multicolumn{2}{c}{$0.34\pm 0.07$} \\
Ne  &  \multicolumn{2}{c}{$0.28\pm 0.09$} \\
Mg  &  \multicolumn{2}{c}{$0.8^{+0.6}_{-0.2}$} \\
Si  & \multicolumn{2}{c}{$9.1\pm 1.5$}\\
S   & \multicolumn{2}{c}{$5 \pm 2$}\\
Fe  & \multicolumn{2}{c}{$4 \pm 2$}\\

\noalign{\smallskip}
\NH ($10^{20}$~cm$^{-2}$)     & \multicolumn{2}{c}{$8.8\pm 0.5$}\\
\noalign{\smallskip}
$\chi^2/\nu$                  &  
\multicolumn{2}{c}{1678/1285} \\
                        \hline\noalign{\smallskip}    
                \end{tabular}
        \end{flushleft}
\end{table}

\section{Spectral fitting}
Fitting six spectral regions simultaneously has the obvious disadvantage that
the spectral model can become very complex. 
So we took care to constrain the spectral model as far as possible without 
loosing too much of its heuristic qualities.
For each sector we chose to have three to four spectral components. 
A typical configuration consisted of the following components for each 
sky region:
a power law component, one or two  
non-equilibrium ionization (NEI) thermal components and
an absorption component (Morrison \& McCammon \cite{Morrison}).
We only looked for spectral differences in the thermal emission
between the Northern and Southern halves of the remnant.
We assumed uniform abundances for SN 1006.

Based on their deprojection of the ROSAT PSPC image of SN 1006, 
Willingale et al. (\cite{Willingale}) reported that the synchrotron emission
does not seem to originate from all around the remnant, but only from
incomplete shells at the Northwest and Southeast of the remnant. 
Consequently, in our simplest model we assume that no synchrotron emission is
originating from the central region of the remnant.
This model produces a reasonable fit to the LECS spectra, but 
it does not fit the MECS spectra of the central regions above $\sim$4~keV,
where an excess in the observed spectra with respect to the model exists
(Fig.~\ref{hard_excess}). The temperature of the thermal components 
was \Te $\sim 1.5$~keV.
So clearly an additional emission component is needed in order to 
fit the emission from the central regions.
This can be either a thermal component with an higher temperature,
but it could also  mean that the non-thermal emission observed to come
from the rims has in reality cylindrical symmetry, 
in which case the apparent structure of the remnant in X-rays may be
due to an extreme case of limb brightening.
The latter possibility  would be in disagreement with the
above mentioned ROSAT PSPC findings,
but it is conceivable that the synchrotron emission is coming from such 
a thin layer that the deprojection scheme of Willingale et al. 
(\cite{Willingale}) may not have worked adequately.

\begin{figure*}
	\psfig{figure=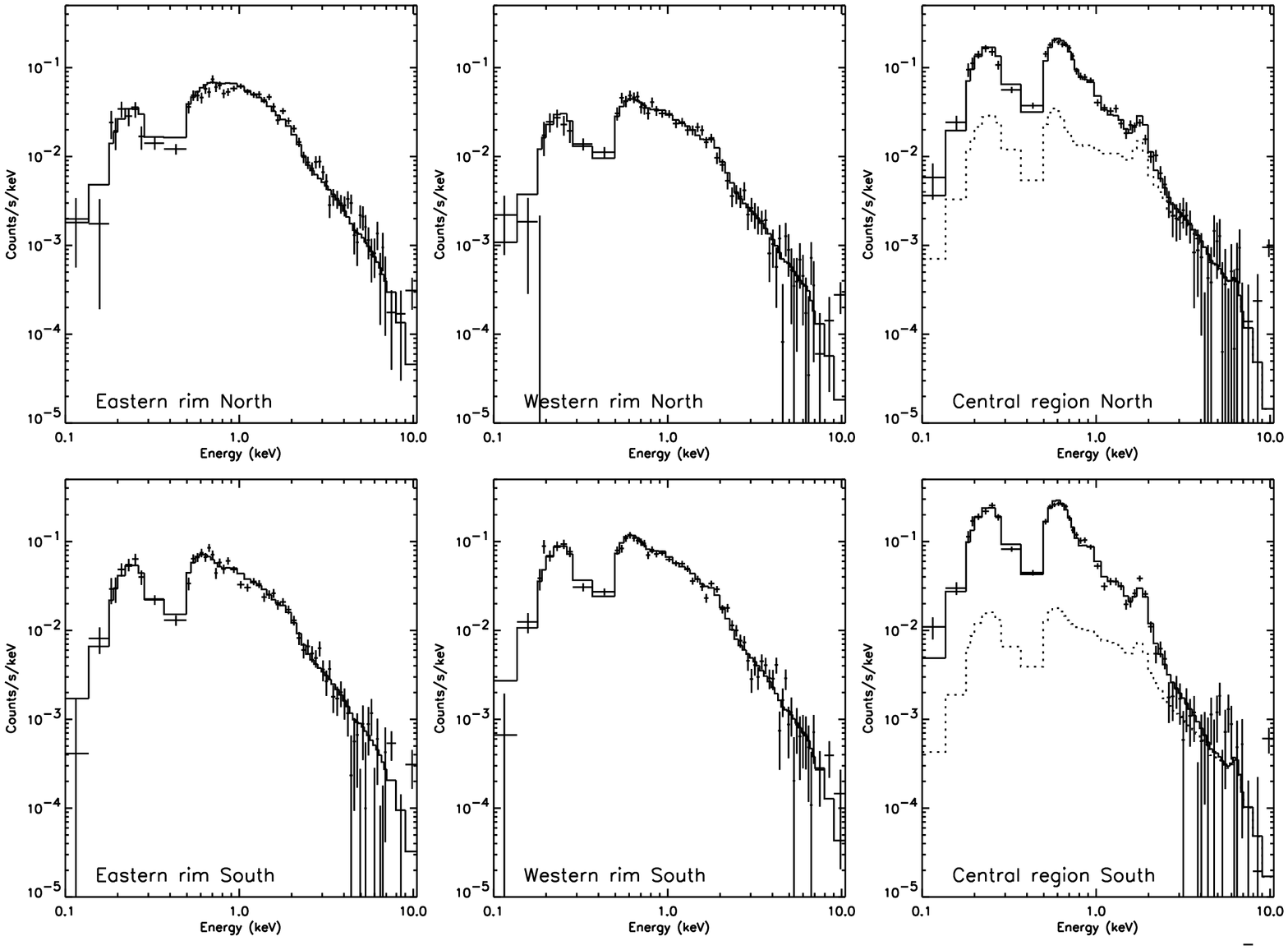,width=16cm,angle=90}
	\psfig{figure=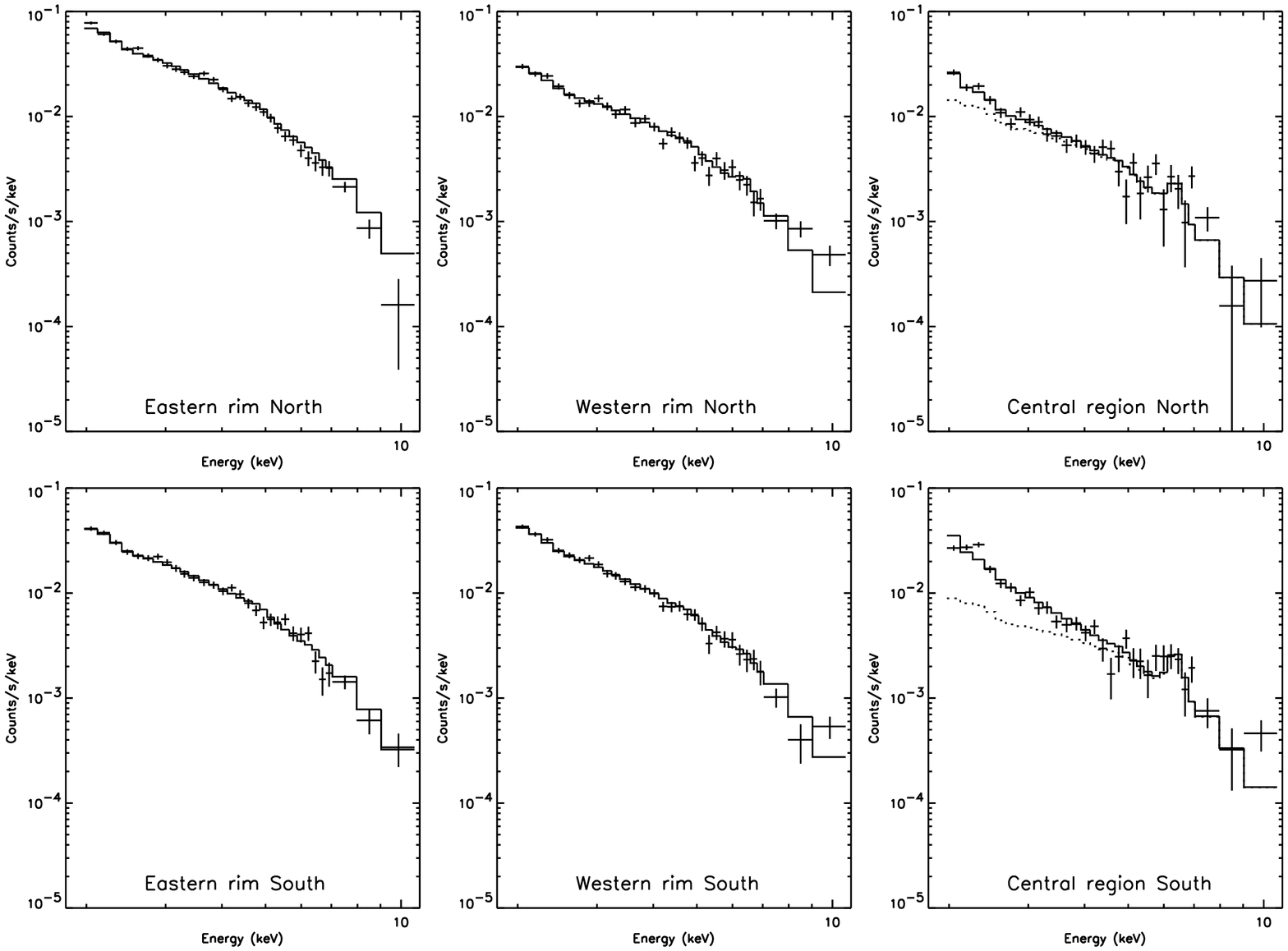,width=16cm,angle=90}
	\caption{The spectra of various regions.
The spectra from different pointings are here summed together for displaying
reasons.
The six upper panels show the spectra as observed by the LECS instrument 
and the six bottom panels are the spectra as observed by the MECS. 
The solid line shows the best fit model convolved with the instrument response
(see Table~\ref{parameters}). The dotted line in the panels with the spectra 
from the central regions shows only the hottest of the two NEI components.
\label{sn1006_spectra}}
\end{figure*}

We investigated both possibilities. 
In the case of an additional power law component we fixed the power law 
index of the central regions to
a value of 2.8, 
similar to the the values found for the rims (see Table~\ref{powerlaw}).
We found, however, that an additional thermal
component offers a better explanation. 
Only a hot thermal component
fits adequately the Fe K emission seen in Fig. \ref{sn1006_spectra}
(most clearly in the right bottom panel). This emission implies that
there is some plasma present with a temperature in excess of $\sim 2$~keV.
An additional power law component has, moreover, the disadvantage that
the result is at a closer look inconsistent: it results in very high 
abundances for the thermal component 
(e.g. 148 times solar for silicon and 356 times solar for iron), 
whereas the emission measure indicated a total mass of $\sim 5$~\msun for
a distance of 1.8~kpc (Laming et al. \cite{Laming96}).
This implies that most of the shock heated plasma is swept 
up material, if SN 1006 was a Type Ia with a mass of $\sim$ 1.4\,\msun,
but then it is not clear why the abundances are so high.
High abundances are also hard to reconcile with the evolved dynamical 
status  of the remnant, which implies that the remnant consists mostly of
swept up interstellar matter 
(Moffet et al. \cite{Moffet}, Long et al. \cite{Long88}).
However, X-ray synchrotron models for supernova remnants 
(Reynolds,\cite{Reynolds98}) show that the synchrotron component
is likely to have a more curved spectrum, which predicts less flux at lower 
energies. 
It may well be that such more advanced models do not have the problems 
mentioned above, although an additional thermal component will still be needed
for producing the Fe K emission.
From a statistical point of view the two models fit the data equally
well, with slightly better fits for the thermal model 
(the model with an additional power law has $\chi^2/\nu = 1689/1287$).

The best fit model with an additional thermal component
(Table~\ref{parameters} and \ref{powerlaw}) is shown in 
Fig.~\ref{sn1006_spectra}.
The total emission measure of
\EM $= (7.4\pm 1.4)\ 10^{62}$~m$^{-3}$kpc$^{-2}$ implies a mass  of 
$M = 8.3\pm 0.8 f_{0.4}^{3/2}d_{1.8}^{3/2}$\msun, 
where $f_{0.4}$ is the volume filling factor divided by 0.4 
(cf. Willingale et al. \cite{Willingale}) and $d_{1.8}$ is 
the distance in units of 1.8~kpc (Laming et al. \cite{Laming96}).
The volume estimate assumes a spherical remnant with a diameter of 30\arcmin.
For this model the implied silicon mass is 0.05\,\msun\ and the iron mass is 
0.06\,\msun, which is much lower than the 0.5\,\msun\ of iron expected in 
remnants 
of Type Ia supernovae. 
The iron abundance reported here is, however, higher than reported by 
Koyama et al. (\cite{Koyama95}).
We do not find significant variations in temperature or 
ionization between the Northern and Southern half of the remnant,
but, as Fig.~\ref{sn1006_spectra} shows, 
the hottest component seems more dominant in the Northern region.
An interesting feature in the spectra of the central regions is the
iron K-shell emission around 6.4~keV.
This is especially apparent in the lower right panel of 
Fig. \ref{sn1006_spectra}.
The centroid of this emission 
($6.3\pm 0.2$~keV)
is consistent with emission from iron in low ionization stages 
(caused by inner shell ionizations and excitations).
In the next section we will discuss the very low value of the
ionization parameter (\net).

\begin{table}
	\caption{The best fit parameters for the non-thermal emission of SN 1006.\label{powerlaw}}
	\begin{flushleft}          
                \begin{tabular}{lll}
                        \hline\noalign{\smallskip}
  & norm & index\\
  & (s$^{-1}$m$^{-2}$keV$^{-1}$ @ 1~keV) &\\
			\noalign{\smallskip}
                        \hline\noalign{\smallskip}
Northeast &  $181\pm 10$ & $2.81 \pm 0.05$\\
Southeast &  $83 \pm 10$  & $2.76 \pm 0.08$\\
Northwest &  $86 \pm 13$  & $2.90 \pm 0.16$\\
Southwest &  $76 \pm 9$  & $2.83 \pm 0.12$\\
                        \noalign{\smallskip}\hline
		\end{tabular}
	\end{flushleft}
\end{table}

Our best fit absorption column is a factor two 
lower than the value found by Koyama et al. (\cite{Koyama95}), 
but is higher than the value reported by Willingale et al. 
(\cite{Willingale}). 
The higher ASCA value is not very surprising, because 
ASCA was less reliably calibrated below $\sim 1$~keV and ASCA is not sensitive
below 0.4~keV. The fact that the ROSAT PSPC and BeppoSAX LECS show that
there is emission at 0.2~keV is, however, 
a clear indication that the absorption column is not as high as indicated by
the ASCA spectrum.
We think that our absorption estimate is consistent with the 
ROSAT PSPC spectra if one 
takes into account that we use non-equilibrium ionization models and in this
case such a model produces more line radiation below 1~keV than an 
equilibrium 
model, which implies a higher absorption to explain the flux levels. 

The power law index of the non-thermal spectra are somewhat lower than 
indicated by the ASCA spectra and higher than found for the ROSAT PSPC data 
(see Table~\ref{powerlaw}).
We checked for gradual steepening of the spectrum, which is expected on 
theoretical grounds (Reynolds \cite{Reynolds98}).
Indeed, we found some evidence that on average the power law index changes 
from $2.65 \pm 0.21$  below 2~keV to $2.81\pm 0.05$ above 2~keV.
However, the scatter in the four points is rather large and we can, 
furthermore, not exclude that the difference in slope is due to calibration 
uncertainties, as there are still problems with the intercalibration 
of the LECS and MECS instruments.

\section{Interpretation}
Our fits indicate that the spectra of the central 
region are best fitted with two NEI components with a very low 
value for the ionization parameter.
The very low ionization is not surprising as previous studies
indicate that the density of the ISM surrounding the remnant is very
low: estimates of the pre-shock hydrogen number density vary from
0.04~cm$^{-3}$ (Laming et al. \cite{Laming96}) to
1~cm$^{-3}$ (Winkler \& Long \cite{Winkler}).
Our mass estimate implies a pre-shock density of 0.1~cm$^{-3}$.
For the Sedov model the average ionization parameter (\net) is approximately
$n_0t$, where $n_0$ is the pre-shock hydrogen density and 
$t$ the age of the remnant.
So for the shocked interstellar gas we estimate 
\net$\sim 3\, 10^{15}$~m$^{-3}$s.
For the ionization parameter of the shocked ejecta we can find an upper 
limit by assuming that 1.4\,\msun\ of ejecta is completely ionized and that 
it has a volume filling factor of 0.4.
This gives an electron density of 0.04~cm$^{-3}$ and implies
\net $\simlt 1.2\, 10^{15}$~m$^{-3}$s. 
The \net\ value found by fitting the X-ray spectra agrees with
our current understanding of SN 1006.

The low ionization has, however, some intriguing aspects.
For example, our models indicate that most of the silicon is in the form
of Si~V to Si~XII. So with the silicon lines around
1.8~keV we are only observing the top of the iceberg.
However, the silicon L-shell emission contributes significantly to the flux
between 0.1~keV and 0.4~keV. 
Also L-shell emission from magnesium, sulphur and argon, 
and K-shell emission from carbon contribute to the flux below 0.5~keV.
Unfortunately, L-shell emission is notoriously complicated;
only the iron L-shell emission has been investigated in substantial detail.
The LECS spectral resolution below 0.5~keV is not good enough to 
see possible discrepancies in the atomic data, but the resulting
uncertainties in the flux below 0.5~keV may have affected the spectral
modeling.
This demonstrates the significance of the broad spectral range covered by
the two BeppoSAX instruments.
The observation of emission from iron around 6.4~keV illustrates
that even ions in low ionization stages emit some X-ray line emission, 
because of inner shell ionizations and excitations
(in this respect SN1006 seems to be similar to RCW 86, 
see Vink et al. \cite{Vink97}).
It would be interesting to search for similar emission features
from Ar and Ca with future X-ray detectors.

The apparent two temperature structure is not surprising.
SN 1006 is a large remnant and it would be more surprising if the whole 
remnant could be characterized by a single temperature.
Furthermore, there are several possible explanations for the two temperature
structure.
Hydrodynamical shock models, such as the Sedov model
(Sedov~\cite{Sedov}) predict temperature gradients. 
Models including a reverse shock (e.g. Chevalier \cite{Chev82})
even predict a two temperature regime. The lowest temperature
is associated with the ejecta heated by the reverse shock, whereas the
highest temperatures are attained by the shocked interstellar medium.
However, hydrodynamical models predict the ion temperature,
whereas the shape of the thermal X-ray continuum is determined by the electron 
temperature, which, in absence of electron-ion equilibration, 
can in principle be as much as a factor 
1835 (the electron/proton mass ratio) lower than the ion temperature. 
It can be even more if the plasma consist of highly enriched supernova ejecta.
In these extreme cases, however, we would not observe thermal X-ray 
emission at all.
The equilibration process is still poorly understood. 
In the case of SN 1006 there is strong evidence for poor 
equilibration of the electron and ion temperature 
(Laming et al. \cite{Laming96}). 
The evidence comes from UV spectroscopy of the Northern shock front, 
but the equilibration may very well vary over the remnant.
Another possible cause for temperature variations
over the remnant are density differences.
There is some evidence that SN 1006 has a very strong density gradient
from the front to the  back of the remnant
(Hamilton et al. \cite{Hamilton97}).
Yet another interesting possibility is that the shock heating process does not 
produce a Maxwellian electron distribution, since the low density in SN 1006 
may prevent a rapid thermalization of the electron distribution
(Laming \cite{Laming98}, see also Itoh~\cite{Itoh}, 
Hamilton \& Sarazin \cite{Hamilton84}).
The hot component could then be associated with bremsstrahlung from the tail 
of the non-thermal electron distribution.

For practical reasons our fitting model has been largely a phenomenological 
model.
It is important to realize this when interpreting the best fit parameters.
For instance, our use of single/double 
NEI components does not account for the fact that there are temperature and 
ionization gradients. Or to be more precise: we suppose that these 
gradients can be approximated by two thermal plasma components.
On the other hand models incorporating temperature and ionization gradients,
such as spectral models based on the Sedov model 
(e.g. Kaastra \& Jansen \cite{Kaastra93}), 
suppose that the gradients are well defined, whereas in reality
the gradients may very well be more complicated, because of the
pre-supernova density structure of the interstellar medium or because of
the non-equilibration of ion and electron temperature, as discussed above.
In our opinion a more severe problem with our modeling is
that we use uniform abundances for the whole remnant, which is probably not 
correct, as there is very likely freshly shocked ejecta present consisting
entirely of metals, on the one hand, and shocked interstellar gas with solar
abundances, on the other hand. 
This might for example affect our total mass estimate, which may be lower 
because a pure metal plasma is an efficient bremsstrahlung emitter.
However, for more realistic models we need either data from more sensitive 
X-ray instruments, or  we need a clear understanding of the thermal and 
abundance structure of the remnant.
For instance, with future X-ray missions we may be able to identify 
spatially or spectrally an ejecta emission component and observe the 
(projected) temperature structure of the remnant directly.

The overall structure of SN 1006, as emerging from other 
observations, appears complicated and is not very well understood.
Some observations seem even contradictory.
For example, 
UV absorption measurements indicate that a lot of the ejecta are 
still in free expansion (Hamilton et al. \cite{Hamilton97}),
but the kinematics of the remnant, on the other hand, show that 
SN 1006 is dynamically evolved implying that most of the ejecta has been 
shocked by now (Moffet et al. \cite{Moffet}, Long et al. \cite{Long88}).
Finally, it is peculiar that a remnant that has such a high degree of
symmetry with respect to the Southeast/Northwest axis 
(Roger et al. \cite{Roger}) seems to have a strong front-back asymmetry 
(Hamilton et al. \cite{Hamilton97}). 
A strong deviation from cylindrical symmetry has also been reported by
Willingale et al. (\cite{Willingale}).

\section{Conclusion}
Our analysis indicates that a one temperature model does not fit the
thermal spectrum of SN 1006 adequately. 
An additional thermal component with a temperature in excess of 3~keV
is needed.
The alternative, an additional power law component, cannot be excluded, 
but in this case the abundances of the low temperature thermal component
become exceptionally high, which is inconsistent with the idea that SN 1006 
must have swept up a considerable amount of interstellar material.

The values for the ionization parameters for the thermal components appear
to be very low (\net $\simlt 3\ 10^{15}$~m$^{-3}$s), 
but are consistent with the electron densities that we derive from 
the emission measure. 
Note, however, that in reality a range of ionization
values will be present. 
For instance, the Sedov model (see Kaastra \& Jansen \cite{Kaastra93}) 
predicts a superposition of plasma ionization values ranging from roughly 0.2 
in the center of the remnant up to 2 times the average value at a radius of 
0.9 times the shock radius.
Inhomogeneities in the pre-shock interstellar medium may extend this range 
upward and downward.
For such a low ionization parameter the 
L-shell emission of Mg, Si and S contributes substantially to the flux
below 0.5~keV. This explains that our best fit hydrogen absorption column
(\NH $= (8.8\pm 0.5)\ 10^{20}$~cm$^{-2}$)
turns out to be higher than found by Willingale et al. (\cite{Willingale}).

Noteworthy is an iron K shell emission feature in the spectrum. 
The centroid of the line emission is $6.3\pm0.2$~keV, which indicates
that the emission is the result of inner shell ionizations and excitations.
The fitted abundances clearly indicate non-solar abundances with, 
in particular, a very high silicon abundance.
This is in agreement with the analysis of ASCA data 
(Koyama et al. \cite{Koyama95}). 
We did not find  significant differences in the temperatures and 
ionization parameters for the spectra of the Northwestern and Southeastern
halves of the remnant, but the hottest component seems more dominant in the 
Northern region.

Observations in the near future with Chandra (AXAF), XMM and Astro E
will be able to reveal new information on the remnant which is needed in order
to obtain a consistent model for this remnant.
In particular, a spatial or a spectral separation of emission originating 
from shocked ejecta and shocked interstellar gas will be very useful 
for understanding more about the structure and abundances of 
this interesting but rather enigmatic supernova remnant.

\begin{acknowledgement}
We thank Martin Laming for stimulating discussions on SN 1006.
This research has made use of data obtained through the High Energy
Astrophysics Science Archive Research Center Online Service, provided
by the NASA/Goddard Space Flight Center. This work was financially
supported by NWO, the Netherlands Organization for Scientific
Research.                         
\end{acknowledgement}

\end{document}